\documentstyle[12pt]{article}
\begin{document}
\topmargin=-1.0cm
\evensidemargin=0cm
\oddsidemargin=0cm
\newcommand{\tilphi}{\tilde{\Phi}}

\newcommand{\bea}[1]{\begin{eqnarray} #1 \end{eqnarray}}

\newcommand{\deri}[2]{\frac{\partial #1}{\partial #2}}

\newcommand{\bra}[1]{\bigl\langle #1 \big|}

\newcommand{\ket}[1]{\big| #1 \bigr\rangle}

\newcommand{\bket}[2]{\bigl\langle #1 \big| #2 \bigr\rangle}

\newcommand{\intinf}{\int^{\infty}_{-\infty}}

\newcommand{\dpi}[2]{\frac{d^#1 #2}{(2 \pi)^#1}}

\newcommand{\spro}[2]{\vec{#1}\cdot\vec{#2}}

\newcommand{\kakko}[1]{\left(#1\right)}

\newcommand{\tkakko}[1]{\left\{#1\right\}}

\newcommand{\dkakko}[1]{\left[#1\right]}

\newcommand{\bee}[1]{\begin{equation}#1\end{equation}}

\newcommand{\lag}{{\cal L}}

\newcommand{\expect}[1]{\left\langle #1 \right\rangle}

\newcommand{\nonum}{\nonumber\\}

\newcommand{\ry}{{\rm Ry}}

\newcommand{\brkt}[3]
{\left\langle #1 \left| #2 \right| #3 \right\rangle}

\newcommand{\e}{{\rm e}}

\newcommand{\Tr}{{\rm Tr}}
\date{}
\baselineskip=0.6cm
\renewcommand{\appendix}{\renewcommand{\thesection}
{\Alph{section}}\renewcommand{\theequation}
{\Alph{section}.\arabic{equation}}
\setcounter{equation}{0}\setcounter{section}{0}}
\begin{titlepage}
\begin{flushright}
{KOBE-TH-00-03}
\end{flushright}
\vspace{1cm}
\begin{center}
{\LARGE Novel Phase Structure of}\\
\vspace{5mm}
{\LARGE Twisted $O(N)$ $\phi^4$ Model on $M^{D-1}\otimes S^1$ }\\
\vskip1.0truein
{\large Katsuhiko Ohnishi}$^{(a)}$
\footnote{E-mail: {\tt katuhiko@oct.phys.sci.kobe-u.ac.jp}}and
{\large Makoto Sakamoto}$^{(b)}$
\footnote{E-mail: {\tt sakamoto@phys.sci.kobe-u.ac.jp}}
\vskip0.2truein
$^{(a)}$ {\it Graduate School of Science and Technology,\\
Kobe University, Rokkodai, Nada, Kobe 657-8501, Japan\\
\vspace*{2mm}
$^{(b)}$Department of Physics,\\
Kobe University, Rokkodai, Nada, Kobe 657-8501, Japan }
\end{center}
\vskip0.5truein \centerline{\bf Abstract} \vskip0.13truein
We study the $O(N)$ $\phi^4$ model compactified 
on $M^{D-1}\otimes S^1$,
which allows to impose twisted boundary conditions 
for the $S^1$-direction.
The $O(N)$ symmetry can be broken to $H$ explicitly 
by the boundary conditions and 
further broken to $I$ spontaneously by vacuum expectation values 
of the fields.
The symmetries $H$ and $I$ are completely classified and the model
turns out to have unexpectedly a rich phase structure.
The unbroken symmetry $I$ is shown to depend 
on not only the boundary conditions
but also the radius of $S^1$,
and the symmetry breaking patterns are found to be unconventional.
The spontaneous breakdown of the translational invariance 
is also discussed.

\end{titlepage}
\newpage
\baselineskip 20 pt
\pagestyle{plain}
\vskip0.2truein
\setcounter{equation}{0}
\vskip0.2truein
\section{Introduction}
Recently,
higher dimensional theories with extra dimensions have 
revived and
have vastly been discussed from various points of view 
\cite{LED,RS,DS}.
In many such scenarios,
nontrivial backgrounds,
such as magnetic flux,
vortices,
domain walls and branes,
turn out to be a key ingredient.
It would be of great importance to study physical consequences 
caused by
the nontrivial backgrounds thoroughly.

In this letter,
we shall concentrate on a simple situation that \lq\lq magnetic'' flux
passes through a circle $S^1$.
Physically,
this system may equivalently be described by the system 
without flux but with fields
obeying twisted boundary conditions for the $S^1$-direction.
In the following,
we shall take the latter point of view for a technical reason.
Even though the situation we consider is very simple,
physical consequences caused by boundary conditions turn out to be 
unexpectedly rich,
as we will see later.
The parameter space of a model on $M^{D-1}\otimes S^1$ is,
in general,
wider than that of the model on $M^D$,
and is spanned by twist parameters specifying boundary conditions 
\cite{Isham,Hosotani},
in addition to parameters appearing in the action.
One of characteristic features of such models is the appearance of 
a critical radius of $S^1$,
at which some of symmetries are
broken/restored.
The spontaneous breakdown of the translational invariance 
for the $S^1$-direction is another characteristic feature 
\cite{STT1,STT2}.

The paper is organized as follows:
In the next section,
we discuss a general feature of scalar field theories 
on $M^{D-1}\otimes S^1$
and allowed boundary conditions.
In Section 3,
the $O(N)$ $\phi^4$ model on $M^{D-1}\otimes S^1$ 
with the antiperiodic boundary condition is studied.
In Section 4,
general twisted boundary conditions are investigated,
and the spontaneous symmetry breaking caused 
by nonvanishing vacuum
expectation values is classified.
In Section 5,
the model is reanalyzed from a $(D-1)$-dimensional field
theory point of view.
Section 6 is devoted to conclusions and discussions.

\section{A General Discussion}
In this section,
we shall discuss a general feature of 
scalar field theories compactified 
on $M^{D-1}\otimes S^1$.
Let us consider an action which consists 
of $N$ real scalar fields 
$\phi_i$ $(i=1,\cdots,N)$\footnote{Repeated indices are 
generally summed, 
unless otherwise indicated.}
\bee{
	S
=
	\int d^{D-1}x\int^{2\pi R}_0dy
	\tkakko{
		-\frac{1}{2}\partial_A\phi_i(x^\nu,y)\partial^A\phi_i(x^\nu,y)
		-V(\phi)
	}\ ,
\label{1}
}
where the index $A$ runs from $0$ to $D-1$,
and $x^\nu$ $(\nu=0,\cdots,D-2)$ and $y$ are 
the coordinates on $M^{D-1}$ and $S^1$,
respectively.
The radius of $S^1$ is denoted by $R$.
Suppose that the action has a symmetry $G$.
Since $S^1$ is multiply-connected,
we can impose a twisted boundary condition on $\phi_i$
\cite{Isham,Hosotani} such as 
\bee{
	\phi_i(x^\nu,y+2\pi R)
=
	U_{ij}\phi_j(x^\nu,y)\ .
\label{2}
}
The matrix $U$ must belong to $G$,
otherwise the action would not be single-valued.
If $U$ is not proportional to the identity matrix,
the symmetry group $G$ will be broken to its subgroup $H$,
which consists of all the elements of $G$ commuting with $U$.
Note that this symmetry breaking caused by the boundary condition 
is not 
spontaneous but explicit.

In order to find the vacuum configuration of $\phi_i(x^\nu,y)$,
one might try to minimize the potential $V(\phi)$.
This would,
however,
lead to wrong vacua in the present model \cite{STT1,STT2}.
To find the true vacuum configuration,
it is important to take account of the kinetic term 
in addition to the potential term.
This is because the translational invariance could be broken 
and the vacuum configuration might be 
coordinate-dependent.
Thus,
the vacuum configuration will be obtained by solving 
a minimization problem of the following functional:
\bee{
	{\cal E}[\phi,R]
\equiv
	\int^{2\pi R}_0dy
		\tkakko{
			\frac{1}{2}\kakko{\frac{d\phi_i(y)}{dy}}^2
		+
			V(\phi)
		}\ ,
\label{3}
}
where we have assumed that the translational invariance of
the uncompactified $(D-1)$-dimensional 
Minkowski space-time is unbroken.\footnote{This is true, at least, 
at the classical level.}

In general,
solving the minimization problem may not be an easy task 
because we must minimize the functional ${\cal E}[\phi,R]$
with the boundary condition (\ref{2}).
Although we have no general procedure to solve 
the minimization problem,
we can present candidates of the vacuum configuration 
of $\phi_i(y)$ for 
some class of twisted boundary conditions.
Suppose that $G$ is a continuous symmetry and that 
the matrix $U$ in Eq.(\ref{2})
can be expressed as $U=e^X$,
where $X$ belongs to the algebra of $G$.
($U$ should continuously be connected to the identity in $G$.)
Then, a candidate of the vacuum configuration will be given by
\bee{
	\bar{\phi}_i(y)
=
	(e^{\frac{y}{2\pi R}X})_{ij}v_j\ ,
\label{4}
}
where $v_i$ $(i=1,\cdots,N)$ are constants.
Note that $\bar{\phi}_i(y)$ satisfy the desired 
boundary condition (\ref{2}).
Even if $U$ cannot continuously be connected to the identity in $G$,
we could find a configuration such as Eq.(\ref{4}) by restricting
some of $v_i$ to zero.
In fact,
we will see later that the vacuum configuration 
can be written into the form (\ref{4}) 
in the $O(N)$ $\phi^4$ model (except for $N=1)$.

\section{$O(N)$ $\phi^4$ Model with the Antiperiodic 
Boundary Condition}
We shall now investigate the $O(N)$ $\phi^4$ model 
whose potential is given by
\bee{
	V(\phi)
=
	-\frac{\mu^2}{2}\phi_i\phi_i+\frac{\lambda}{8}\kakko{\phi_i\phi_i}^2\ .
\label{5}
}
Since the phase structure is trivial for a positive squared mass,
we will assume $\mu^2>0$ in the following analysis.
The boundary condition for $\phi_i(y)$ is taken to be antiperiodic, i.e.
\bee{
	\phi_i(y+2\pi R)
=
	-\phi_i(y)
\qquad \mbox{for $i=1,\cdots,N$\ .}
\label{6}
}
General twisted boundary conditions will be discussed in the next section.
Since $U=-{\bf 1}$,
the twisted boundary condition (\ref{6}) does not break the $O(N)$ symmetry,
and hence the unbroken symmetry $H$,
which is consistent with the boundary condition,
is $O(N)$ itself.

Let us first consider the case of even $N$.
In this case,
it may be convenient to introduce the $N/2$ complex fields by
\bee{
	\Phi_a(y)
\equiv
	\frac{e^{-i\frac{y}{2R}}}{\sqrt{2}}
		\kakko{
			\phi_{2a-1}(y)+i\phi_{2a}(y)
		}
\qquad
	\mbox{for $a=1,\cdots,\frac{N}{2}$\ .}
\label{7}
}
It should be noticed that $\Phi_a(y)$ obey the periodic
boundary condition, i.e.
\bee{
	\Phi_a(y+2\pi R)
=
	+\Phi_a(y)
\qquad
\mbox{for $a=1,\cdots,\frac{N}{2}$}\ .
\label{9}
}
Inserting Eq.(\ref{9}) into ${\cal E}[\phi,R]$,
we may write
\bee{
	{\cal E} [\phi,R]
=
	{\cal E}^{(1)}[\Phi,R]
	+
	{\cal E}^{(2)}[\Phi,R]\ ,
\label{10}
}
where
\bea{
	{\cal E}^{(1)}[\Phi,R]
&\equiv&
	\int^{2\pi R}_0dy
		\tkakko{
			\Big|\frac{d\Phi_a}{dy}\Big|^2
			-
			\frac{i}{2R}
				\kakko{
					\Phi^\ast_a\frac{d\Phi_a}{dy}
					-
					\frac{d\Phi^\ast_a}{dy}\Phi_a
				}
		}\ ,
\label{11}
\\
	{\cal E}^{(2)}[\Phi,R]
&\equiv&
	\int^{2\pi R}_0dy
		\tkakko{
			\kakko{\frac{1}{4R^2}-\mu^2}\big|\Phi_a\big|^2
			+
			\frac{\lambda}{2}\kakko{\big|\Phi_a\big|^2}^2
		}\ .
\label{12}
}
Our strategy to find the vacuum configuration,
which minimizes the functional (\ref{10}),
is as follows:
We shall first look for configurations which minimize each of 
${\cal E}^{(1)}[\Phi,R]$ and ${\cal E}^{(2)}[\Phi,R]$,
and then construct configurations which minimize both of 
them simultaneously.

Let us first look for configurations which minimize 
${\cal E}^{(1)}[\Phi,R]$.
To this end,
we may expand $\Phi_a(y)$ in the Fourier-series,
according to the boundary condition (\ref{9}),
as
\bee{
	\Phi_a(y)
=
	\sum_{n\in {\bf Z}}\varphi_a^{(n)}e^{i\frac{n}{R}y}
\qquad
	\mbox{for $a=1,\cdots,\frac{N}{2}$}\ .
\label{13}
}
Inserting Eq.(\ref{13}) into ${\cal E}^{(1)}[\Phi,R]$, we find
\bee{
	{\cal E}^{(1)}[\Phi,R]
=
	\frac{2\pi}{ R}\sum_{n\in {\bf Z}}
		\dkakko{
			\kakko{n+\frac{1}{2}}^2-\kakko{\frac{1}{2}}^2
		}
	\big|\varphi_a^{(n)}\big|^2\ .
\label{14}
}
Since $(n+1/2)^2-(1/2)^2\ge 0$ for all $n\in {\bf Z}$,
${\cal E}^{(1)}[\Phi,R]$ is positive semi-definite.
The configuration which gives ${\cal E}^{(1)}[\Phi,R]=0$ is
found to be of the form
$\Phi_a(y)=\varphi_a^{(0)}+\varphi_a^{(-1)}e^{-i\frac{y}{R}}
\ (a=1,\cdots,N/2)$,
where $\varphi^{(0)}_a$ and $\varphi^{(-1)}_a$ are arbitrary 
complex constants.
Let us next look for configurations 
which minimize ${\cal E}^{(2)}[\Phi,R]$.
We find that the configuration which minimizes 
${\cal E}^{(2)}[\Phi,R]$ is
$\Phi_a(y)=0$ for $R\le 1/(2\mu)$
and
$\big|\Phi_a(y)\big|^2=(\mu^2-1/(2R)^2)/\lambda$ for $R>1/(2\mu)$.
Combining the above two results
and performing an appropriate orthogonal $O(N)$ transformation,
we conclude that in terms of $\phi_i$ 
the vacuum configuration,
which minimizes both of ${\cal E}^{(1)}[\Phi,R]$
and ${\cal E}^{(2)}[\Phi,R]$ simultaneously,
can take to be of the form
\bee{
	\expect{\phi_i(x^\nu,y)}
=
			\left\{
			\begin{array}{ll}
			(0,0,\cdots,0)		&	\mbox{for $R\le\frac{1}{2\mu}$} \\
			(v\cos(\frac{y}{2R}),v\sin(\frac{y}{2R}),0,\cdots,0)&	
			\mbox{for $R>\frac{1}{2\mu}$\ ,}
			\end{array}
		\right.
\label{18}
}
where $v=\sqrt{2(\mu^2-1/(2R)^2)/\lambda }$.
It follows that for $R\le 1/(2\mu)$ the $O(N)$ symmetry is unbroken,
while for $R>1/(2\mu)$ the spontaneous symmetry breaking occurs and
the $O(N)$ symmetry is broken to $O(N-2)$.
It is interesting to contrast this result with that 
of the $O(N)$ $\phi^4$ model
with the periodic boundary condition,
for which the $O(N)$ symmetry is spontaneously broken 
to $O(N-1)$ irrespective of $R$.

We now proceed to the case of odd $N$.
In this case,
we cannot apply the same method, as was done above,
to find the vacuum configuration
because we cannot take a complex basis 
such as Eq.(\ref{7}) for odd $N$ and
because the twist matrix $U=-{\bf1}$ cannot continuously 
be connected to 
the identity matrix.
Nevertheless,
we can show that the problem to find the vacuum configuration 
for odd $N$ reduces to
that for even $N$ (expect for $N=1$).
The trick is to add an additional real field $\phi_{N+1}(y)$ 
satisfying the antiperiodic boundary condition to the action 
in order to form 
the $O(N+1)$ $\phi^4$ model.
It follows from the previous analysis that the vacuum configuration 
will be found to be of 
the form (\ref{18}) since $N+1$ now becomes an even integer.
The fact that the configuration space spanned by  
$\{\phi_i(y)$, $i=1,\cdots,N+1\}$ contains that by 
$\{\phi_i(y)$, $i=1,\cdots,N\}$
implies that the vacuum for odd $N$ is also given by Eq.(\ref{18}),
and hence the spontaneous symmetry breaking from 
$O(N)$ to $O(N-2)$ can occur for $R>1/(2\mu)$.
The exception is the model with $N=1$.
In this case,
there is no continuous symmetry and the $O(1)$ model has only a 
discrete symmetry of $G=H=Z_2$.
The $O(1)$ $\phi^4$ model has been investigated in Ref.\cite{STT1}
and the vacuum configuration has been found to be
\bee{
	\expect{\phi (x^\nu,y)}
=
			\left\{
			\begin{array}{ll}
				0		&	\mbox{for $R\le\frac{1}{2\mu}$} \\
				\frac{2k\mu}{\sqrt{\lambda(1+k^2)}}\ 
				{\rm sn} \kakko{\frac{\mu}{\sqrt{1+k^2}}(y-y_0),k}
						&	\mbox{for $R>\frac{1}{2\mu}$\ .}
			\end{array}
		\right.
\label{19}
}
Here,
${\rm sn}(u,k)$ is the Jacobi elliptic function whose period is $4K(k)$,
where $K(k)$ denotes the complete elliptic function of the first kind.
The $y_0$ is an integration constant 
and the parameter $k$ $(0\le k <1)$ is determined by the relation
$\pi R \mu =\sqrt{1+k^2}K(k)$.
Thus,
the $Z_2$ symmetry is unbroken for $R\le 1/(2\mu)$,
while it is broken spontaneously for $R>1/(2\mu)$.

\section{General Twisted Boundary Conditions}
In this section,
we shall construct the vacuum configurations 
of the $O(N)$ $\phi^4$ model on $M^{D-1}\otimes S^1$
for general twisted boundary conditions and clarify the phase structure.

To discuss general boundary conditions,
it is convenient to transform the matrix $U$ in Eq.(\ref{2})
by means of a real orthogonal transformation into the normal form.
It is known that any matrix $U$ belonging to $O(N)$ can be transformed,
by an orthogonal transformation,
into a block diagonal form whose diagonal elements are 
one of $1$, $-1$ and
a two dimensional rotation matrix \cite{Schutz}.
In this basis,
we may arrange the boundary conditions for $\phi_i(y)$ as follows:
\bea{
	\phi_{a}^{(\alpha_0)}(y+2\pi R)
&=&
	+\phi_{a}^{(\alpha_0)}(y)
\qquad
	\mbox{for $a=1,\cdots,L_0$\ ,}\nonumber\\
\pmatrix{
		\phi_{2b_k-1}^{(\alpha_k)}(y+2\pi R)	\cr
		\phi_{2b_k}^{(\alpha_k)}(y+2\pi R)
	}
&=&
	\pmatrix{
		\cos(2\pi\alpha_k)		&		-\sin(2\pi\alpha_k)	\cr
		\sin(2\pi\alpha_k)		&		\cos(2\pi\alpha_k)		
	}
	\pmatrix{
		\phi_{2b_k-1}^{(\alpha_k)}(y)	\cr
		\phi_{2b_k}^{(\alpha_k)}(y)
	}\nonumber\\
& & \qquad 
    \mbox{for $b_k=1,\cdots,\frac{L_{k}}{2}$ and $k=1,\cdots,M-1$\ ,}
	\nonumber\\
\phi^{(\alpha_M)}_{c}(y+2\pi R)
&=&
	-\phi^{(\alpha_M)}_{c}(y)
\qquad
\mbox{for $c=1,\cdots,L_M$}\ ,
\label{20}
}
where $L_0+L_1+\cdots+L_{M-1}+L_M=N$ and
$0=\alpha_0<\alpha_1<\cdots<\alpha_{M-1}<\alpha_M=1/2$.
The above boundary conditions
explicitly break the $O(N)$ symmetry down to
\bee{
	H
=
	O(L_0)\times U({\textstyle \frac{L_1}{2}})\times\cdots
	\times U({\textstyle \frac{L_{M-1}}{2}})\times O(L_M)
\label{21}
}
which is the subgroup of $O(N)$ commuting with the twist matrix $U$.


Let us first consider the case of $L_0\neq 0$.
In this case,
$\phi_{a}^{(\alpha_0)}(y)$ $(a=1,\cdots,L_0)$
satisfy the periodic boundary condition.
Then,
it is easy to show that the vacuum configuration can,
without loss of generality,
be taken into the form
\bee{
	\expect{\phi_1^{(\alpha_0)}(x^{\nu},y)}
=
	\sqrt{\frac{2}{\lambda}}\mu\ ,
\label{22}
}
and other fields vanish.
Thus,
the symmetry $H$ in Eq.(\ref{21}) is
spontaneously broken to
\footnote{For $L_0=1$,
$O(L_0=1)$ means $Z_2$ and the $Z_2$ symmetry is broken completely.}
\bee{
	I=O(L_0-1)\times U({\textstyle \frac{L_1}{2}})\times\cdots
	\times U({\textstyle \frac{L_{M-1}}{2}})\times O(L_M)\ ,
\label{23}
}
irrespective of the value of the radius $R$.


Let us next consider the case of $L_0=0$ and $N=\mbox{even}$.
It is then convenient to introduce the $N/2$ complex fields as
\bee{
	\Phi_{b_l}^{(\alpha_l)}(y)
	\equiv
	\frac{e^{-i\frac{\alpha_l}{R}y}}{\sqrt{2}}
	\kakko{\phi_{2b_l-1}^{(\alpha_l)}(y)+i\phi_{2b_l}^{(\alpha_l)}(y)}
\quad
	\mbox{for $b_l=1,\cdots,\frac{L_{l}}{2}$ and $l=1,\cdots,M$}\ .
\label{24}
}
Inserting Eqs.(\ref{24}) into ${\cal E}[\phi,R]$,
we may rewrite it into the form
\bee{
	{\cal E}[\phi,R]
=
	{\cal E}^{(1)}[\Phi,R]+{\cal E}^{(2)}[\Phi,R]+{\cal E}^{(3)}[\Phi,R]\ ,
\label{26}
}
where
\bea{
	{\cal E}^{(1)}[\Phi,R]
&\equiv&
	\int^{2\pi R}_0dy
		\tkakko{
			\Bigg|\frac{d\Phi_{b_l}^{(\alpha_l)}}{dy}\Bigg|^2
			-
			i\frac{\alpha_l}{R}
		\kakko{
		\Phi_{b_l}^{(\alpha_l)\ast}\frac{d\Phi_{b_l}^{(\alpha_l)}}{dy}
		-
		\frac{d\Phi_{b_l}^{(\alpha_l)\ast}}{dy}\Phi_{b_l}^{(\alpha_l)}
				}
		}\ ,\nonumber
\\
	{\cal E}^{(2)}[\Phi,R]
&\equiv&
	\int^{2\pi R}_0dy
		\tkakko{
			\dkakko{\kakko{\frac{\alpha_1}{R}}^2-\mu^2}
			\big|\Phi^{(\alpha_l)}_{b_l}\big|^2
			+
			\frac{\lambda}{2}\kakko{\big|\Phi^{(\alpha_l)}_{b_l}\big|^2}^2
		}\ ,\nonumber
\\
	{\cal E}^{(3)}[\Phi,R]
&\equiv&
	\int^{2\pi R}_0dy
		\dkakko{
			\kakko{\frac{\alpha_l}{R}}^2-\kakko{\frac{\alpha_1}{R}}^2
		}
	\big|\Phi_{b_l}^{(\alpha_l)}\big|^2\ .
\label{29}
}
Since $(\alpha_1)^2<(\alpha_l)^2$ for $l=2,\cdots,M$,
it is not difficult to show that 
in terms of the fields (\ref{24}) the vacuum configuration 
which minimizes every ${\cal E}^{(j)}[\Phi,R]$ $(j=1,2,3)$
simultaneously can,
without loss of generality, 
be taken into the form
\bee{
	\expect{\Phi^{(\alpha_l)}_{b_l}(x^\nu,y)}
=
			\left\{
			\begin{array}{ll}
			0		&	\mbox{for $R\le\frac{\alpha_1}{\mu}$} \\
			\frac{v}{\sqrt{2}}\delta_{\alpha_l,\alpha_1}\delta_{b_l,1}
						&	\mbox{for $R>\frac{\alpha_1}{\mu}$}
			\end{array}
		\right.
\label{30}
}
with $v=\sqrt{2(\mu^2-(\alpha_1/R)^2)/\lambda}$.
It follows that for $R\le\alpha_1/\mu$ 
the symmetry $H$ with $L_0=0$ is unbroken,
while for $R>\alpha_1/\mu$ it is spontaneously broken to 
\footnote{
For $L_1=\cdots=L_{M-1}=0$, the symmetry $H$ is $O(N)$
and is broken to $O(N-2)$ for $R>1/(2\mu)$,
as shown in the previous section.
}
\bee{
	I=U({\textstyle \frac{L_1}{2}}-1)\times U({\textstyle \frac{L_2}{2}})
	\times\cdots\times U({\textstyle \frac{L_{M-1}}{2}})\times O(L_M)\ .
\label{31}
}


Let us finally investigate the case of $L_0=0$ and $N=\mbox{odd}$.
To find the vacuum configuration,
we may perform the trick used in the previous section:
We add an additional real field $\phi_{N+1}(y)$
which satisfies the antiperiodic boundary condition to the action.
Then,
the resulting model may become the $O(N+1)$ model,
which has been analyzed just above since $N+1$ is now even.
The result of the $O(N+1)$ model will tell us that the vacuum 
configuration
for the $O(N)$ model with odd $N$ can be taken into the same form 
as Eq.(\ref{30})
(except for $N=1$).\footnote{
Since for $N=1$ the possible boundary condition is either 
periodic or antiperiodic,
the $O(1)$ model has no new phase more than discussed in the 
previous section.
}
It follows that for $R\le \alpha_1/\mu$ the symmetry $H$ 
with $L_0=0$ is unbroken,
while for $R>\alpha_1/\mu$ the spontaneous symmetry breaking occurs 
and the symmetry $H$ is broken to $I$ given in Eq.(\ref{31}).
\section{Reanalysis with Kaluza-Klein Modes}
In the previous sections,
we have succeeded to reveal the phase structure of the
twisted $O(N)\ \phi^4$ model.
In this section,
we shall reanalyze the model from a $(D-1)$-dimensional
field theory point of view,
and discuss Nambu-Goldstone modes associated with the
broken symmetries and also the symmetry breaking of
the translational invariance for the $S^1$-direction.


To avoid inessential complexities, we shall restrict our considerations
to the case of $L_0, L_M =$ even.
The $N$ real fields (\ref{20}) can then form the $N/2$ complex
fields which are expanded in the Fourier-series as
\bee{
	\frac{1}{\sqrt{2}}
	\kakko{\phi_{2b_l-1}^{(\alpha_l)}(x^\nu,y)
	+i\phi_{2b_l}^{(\alpha_l)}(x^\nu,y)}
	= \sum_{n\in {\bf Z}}
	\varphi_{b_l,n}^{(\alpha_l)}(x^\nu)\ e^{i(\frac{n+\alpha_l}{R})y}
\label{a1}
}
for $l = 0,1,\cdots,M$ and $b_l = 1,2,\cdots,L_l/2$.
Inserting Eq.(\ref{a1}) into Eq.(\ref{3}), we have,
up to the quadratic terms with respect to 
$\varphi_{b_l,n}^{(\alpha_l)}$,
\bee{
	{\cal E}_0[\varphi,R]
	= 2\pi R
	\sum^{M}_{l=0}\sum^{L_l/2}_{b_l=1}\sum_{n\in {\bf Z}}
	m^2_{l,n}\ |\varphi_{b_l,n}^{(\alpha_l)}|^2\ \ ,
\label{a2}
}
where $m^2_{l,n}$ are the squared masses of the Kaluza-Klein
modes $\varphi_{b_l,n}^{(\alpha_l)}$ and are given by
\bee{
	m^2_{l,n} = -\mu^2 + \kakko{\frac{n+\alpha_l}{R}}^2\ \ .
\label{a3}
}
The second term in Eq.(\ref{a3}) is the Kaluza-Klein mass,
which comes from the \lq\lq kinetic" term 
$\frac{1}{2}(\partial_y \phi_i(y))^2$
and which gives a positive contribution to the squared mass term.

For $L_0 \neq 0$, the squared mass $m^2_{0,0}$ for the
modes $\varphi^{(\alpha_0)}_{b_0,0}$ is always negative
irrespective of $R$.
This observation suggests that $\varphi^{(\alpha_0)}_{b_0,0}$
acquire non-vanishing vacuum expectation values,
so that the $O(L_0)$ symmetry is spontaneously broken.
This is consistent with the results obtained in the
previous section.
Taking Eq.(\ref{22}) into account, we should replace the fields
$\varphi_{b_l,n}^{(\alpha_l)}$ by 
$\tilde{\varphi}_{b_l,n}^{(\alpha_l)} +
\frac{\mu}{\sqrt{\lambda}}\delta_{l,0}\delta_{b_l,1}\delta_{n,0}$
and then find that all the squared masses for 
$\tilde{\varphi}_{b_l,n}^{(\alpha_l)}$ become positive
semi-definite, as they should be.
The $L_0 - 1$ massless modes,
${\rm Im}\tilde{\varphi}_{1,0}^{(\alpha_0)}$
and $\tilde{\varphi}_{b_0,0}^{(\alpha_0)}\ (b_0 = 2,3,\cdots,L_0/2)$,
are found to appear and turn out to correspond to the
Nambu-Goldstone modes associated with the broken generators of
$O(L_0)/O(L_0-1)$.

For $L_0=0$, all the squared masses in Eq.(\ref{a3}) are
positive for $R<\alpha_1/\mu$.
The $m^2_{1,0}$ vanishes at $R=\alpha_1/\mu$ and becomes negative
for $R>\alpha_1/\mu$.
This is a signal of the phase transition and is consistent with
the results obtained in the previous section.
Taking Eq.(\ref{30}) into account, we should replace the fields
$\varphi_{b_l,n}^{(\alpha_l)}$ by 
$\tilde{\varphi}_{b_l,n}^{(\alpha_l)}+\frac{v}{\sqrt{2}}\delta_{l,1}
\delta_{b_l,1}\delta_{n,0}$ for $R>\alpha_1/\mu$ and then find
that all the squared masses become positive semi-definite, as they
should be.
The $L_1 - 1$ massless modes,
${\rm Im}\tilde{\varphi}_{1,0}^{(\alpha_1)}$ and 
$\tilde{\varphi}_{b_1,0}^{(\alpha_1)}\ (b_1=2,3,\cdots,L_1/2)$,
are
found to appear and turn out to correspond to the Nambu-Goldstone
modes associated with the broken generators of
$U(\frac{L_1}{2})/U(\frac{L_1}{2}-1)$.
If $L_M=N$, the additional $N-2$ massless modes,
$\tilde{\varphi}_{b_M,-1}^{(\alpha_M)} \ (b_M = 2,3,\cdots,N/2)$,
appear and all the massless modes turn out to form the
Nambu-Goldstone modes associated with the broken generators of
$O(N)/O(N-2)$.

We shall finally discuss the symmetry breaking of the
translational invariance for the $S^1$-direction.
For $L_0\neq 0$, the vacuum expectation values of the fields
are coordinate-independent, so that the translational
invariance  is unbroken.
For $L_0 =0$, the vacuum expectation values depend on the
coordinate $y$ for $R>\alpha_1/\mu$, so that the translational
invariance for the $S^1$-direction is spontaneously broken.
It may be instructive to point out that the translational invariance
for the $S^1$-direction can be reinterpreted as a global $U(1)$
symmetry, which is in fact possessed by the theory after
compactification.
To see this, we note that the translations 
$y\rightarrow y+\epsilon R$ in Eq.(\ref{a1}) can equivalently
be realized by the following $U(1)$ transformations:
\bee{
	U(1):\quad \varphi_{b_l,n}^{(\alpha_l)}
	\longrightarrow e^{i(n+\alpha_l)\epsilon}
\varphi_{b_l,n}^{(\alpha_l)}
\label{a4}
}
from which we may assign a $U(1)$ charge $n+\alpha_l$ to the field
$\varphi_{b_l,n}^{(\alpha_l)}$.
Thus, the spontaneous breakdown of the translational invariance
for the $S^1$-direction may be understood as that of the $U(1)$
symmetry.
For $L_0 \neq 0$, some of $\varphi_{b_0,0}^{(\alpha_0)}$
acquire non-vanishing vacuum expectation values 
but have no $U(1)$ charges, so that the $U(1)$ symmetry is unbroken.
For $L_0 = 0$, some of $\varphi_{b_1,0}^{(\alpha_1)}$ acquire
non-vanishing vacuum expectation values for $R>\alpha_1/\mu$.
Since $\varphi_{b_1,0}^{(\alpha_1)}$ have the nonzero $U(1)$
charge $\alpha_1$, the $U(1)$ symmetry would be broken for
$R>\alpha_1/\mu$.
However, the following modified $U(1)'$ symmetry, which is a
combination of the $U(1)$ symmetry and the $O(N)$ symmetry,
survives as a symmetry even for $R>\alpha_1/\mu$:
\bee{
	U(1)':\quad \varphi_{b_l,n}^{(\alpha_l)}
	\longrightarrow e^{in\epsilon}
\varphi_{b_l,n}^{(\alpha_l)}\ \ .
\label{a4}
}
This is because $\varphi_{b_1,0}^{(\alpha_1)}$ now have
zero $U(1)'$ charge.
Hence, no new Nambu-Goldstone modes are produced other
than those found before.

\section{Conclusions and Discussions}
We have studied the $O(N)$ $\phi^4$ model compactified on
$M^{D-1}\otimes S^1$ with the general twisted boundary conditions.
Since $S^1$ is multiply-connected,
the model can be parametrized by not only the mass and 
the coupling appearing in the action but also the twist matrix appearing
in the boundary condition (\ref{2}).
Thus,
the parameter space of the $O(N)$ model on $M^{D-1}\otimes S^1$ is
much wider than that on $M^D$.
We have succeeded to reveal the rich phase structure and to classify
the patterns of the symmetry breaking/restoration thoroughly.


In this letter,
our analysis has been restricted to the classical level,
and has not taken quantum corrections into account.
When the radius $R$ of $S^1$ is large,
$R$-dependent quantum corrections might be small.
But when $R$ is smaller than the inverse of the mass,
the leading correction to the squared mass turns
out to be proportional to $1/R^2$ for $D=4$ \cite{Quantum}
and hence could drastically change the phase structure 
at the classical level.
Furthermore,
the introduction of gauge fields leads to 
a new interesting feature:
Twisted boundary conditions in the directions of 
the gauge symmetry 
can dynamically be determined through the Hosotani 
mechanism \cite{Hosotani}.
It would be of great interest to analyze $R$-dependent 
quantum corrections
in gauge field theories and the phase structure of symmetries 
systematically.
The work on these subjects will be reported elsewhere.


\vspace{2cm}
\centerline{{ \sc Acknowledgments}}
We would like to thank to H. Hatanaka,
M. Tachibana and K. Takenaga 
for valuable discussions.

\end{document}